\documentclass{article}
\usepackage{spconf,amsmath,graphicx,hyperref}
\usepackage[numbers]{natbib}
\usepackage{graphicx}
\usepackage{tcolorbox}
\usepackage{amsmath} 
\usepackage{url}
\usepackage{multirow}
\usepackage{multicol}
\usepackage{booktabs}
\usepackage{bm}
\usepackage[dvipsnames]{xcolor}
\usepackage{colortbl}
\usepackage{amssymb}
\usepackage{algorithm}
\usepackage{algorithmic}
\usepackage{enumitem}
\usepackage[font=small]{caption} 

\usepackage{CJKutf8}
\usepackage{setspace} 

\title{Direct Simultaneous Translation Activation \\for Large Audio-Language Models}
\name{Pei Zhang$^{*\alpha \beta}$, Yiming Wang$^{*\gamma}$, Jialong Tang$^{\alpha}$, Baosong Yang$^{\alpha}$, Rui Wang$^{\gamma}$,
Derek F. Wong$^{\beta\dagger}$, Fei Huang$^{\alpha}$}

\address{$^{\alpha}$Tongyi Lab, Alibaba Group, $^{\beta}$NLP$^2$CT Lab, University of Macau,\\
$^{\gamma}$School of Computer Science, Shanghai Jiao Tong University}

\begin{document}
\ninept
\maketitle
\begin{abstract}

Simultaneous speech-to-text translation (Simul-S2TT) aims to translate speech into target text in real time, outputting translations while receiving source speech input, rather than waiting for the entire utterance to be spoken. Simul-S2TT research often modifies model architectures to implement read-write strategies. However, with the rise of large audio-language models (LALMs), a key challenge is how to directly activate Simul-S2TT capabilities in base models without additional architectural changes.
In this paper, we introduce {\bf Simul}taneous {\bf S}elf-{\bf A}ugmentation ({\bf SimulSA}), a strategy that utilizes LALMs' inherent capabilities to obtain simultaneous data by randomly truncating speech and constructing partially aligned translation. By incorporating them into offline SFT data, SimulSA effectively bridges the distribution gap between offline translation during pretraining and simultaneous translation during inference.
Experimental results demonstrate that augmenting only about {\bf 1\%} of the simultaneous data, compared to the full offline SFT data, can significantly activate LALMs' Simul-S2TT capabilities without modifications to model architecture or decoding strategy.
\end{abstract}
\section{Introduction}
Simultaneous machine translation (Simul-MT) is an essential research field with broad application in multilingual communication \citep{cho2016can,raffel2017online,arivazhagan2019monotonic,guo2024sillm}. Especially for speech input, Simul-MT enables continuous interpretation to listeners without interrupting the speaker, significantly improving the efficiency of information transmission between cross-lingual communicators \citep{iranzo2020direct,chen2021direct}. For applications like synchronized live subtitles, the latency and accuracy of simultaneous speech-to-text translation (Simul-S2TT) systems are crucial \citep{iranzo2020direct,chen2021direct}.


However, Simul-S2TT faces challenges distinct from offline S2TT. Its main difficulty lies in handling streaming input: the system receives only partial audio segments rather than the full utterance. It must therefore generate translations from incomplete fragments, which may not align with full semantic units or clear linguistic boundaries. This raises the risk of early mistranslations that can propagate and compound, ultimately degrading translation quality.
Existing research primarily tackles these issues through speech segmentation and ``Read-Write (RW)'' strategies \citep{liu2024recent}. Some methods segment speech into complete semantic units for improved recognition \citep{dong2022learning,zhang2023end}, while others introduce new decoder-encoder architectures \citep{zaidi2021decision,liu2021cross}, incremental decoding \citep{polak2023incremental}, or attention constraints \citep{papi2023attention} to adapt offline models to precise RW timing.

\begin{figure}[ht]
  \hfill
  \centering
  \includegraphics[width=\columnwidth]{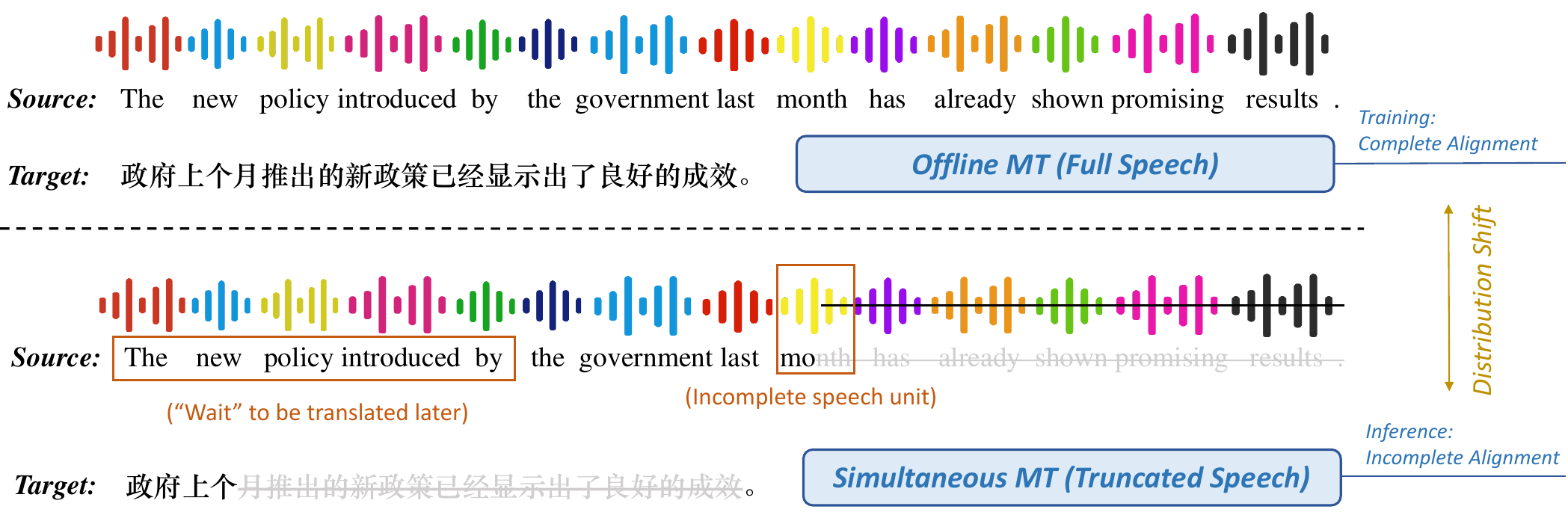}  
  \vspace{-0.3in}
  \caption{Distribution gap in speech-to-text translation: Offline MT fully aligns audio with target text, while simultaneous MT may have partial alignment due to reordering and real-time translation.
}
  \label{img:simul-case}
  \vspace{-0.2in}
\end{figure}

With the rise of Large Audio-Language Models (LALMs) \citep{chu2023qwen,tang2023salmonn,das2024speechverse,chu2024qwen2}, integrating Simul-S2TT capabilities has become essential. Yet, traditional methods often require substantial architectural modifications to support streaming, which hinders the direct deployment of general LALMs in practice.
Therefore, a key unresolved challenge is {\it how to directly activate LALMs' Simul-S2TT capabilities without altering its architecture and decoding strategy}.

We argue that the fundamental difficulty lies in constructing effective training data for truncated speech input paired with its corresponding ground-truth translation output.
To elaborate, the main gap between Simul-MT and Offline-MT ({\it i.e.}, complete end-to-end MT) is {\bf the completeness of information alignment}, as shown in Figure \ref{img:simul-case}.
In Offline-MT, the model fully aligns complete source-side information with the target, ensuring no omissions.
In contrast, when processing truncated speech, Simul-MT often needs to ``wait'' for some speech information to be translated later for handling ``word reordering'' across languages, resulting in incomplete source-to-target alignment.
This alignment difference causes distribution gaps between training and inference. Moreover, truncated speech may contain incomplete semantic units that are also absent during training, further complicating inference for Simul-S2TT tasks.

\begin{figure*}[t]
  \centering
  \vspace{-0.1in}
  \includegraphics[width=1.7\columnwidth]{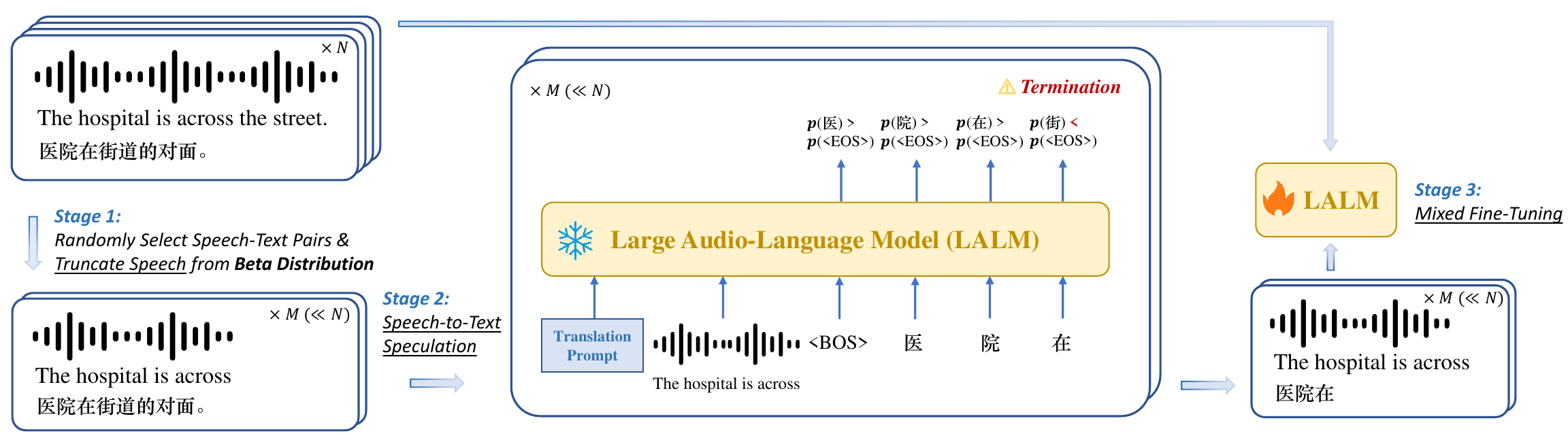}  
  \vspace{-0.12in}
  \caption{The overall pipeline and example of our {\bf Simul}taneous {\bf S}elf-{\bf A}ugmentation ({\bf SimulSA}) method.}
  \label{img:simulsa}
\vspace{-0.1in}
\end{figure*}

Given these challenges, we aim to augment simultaneous data, enabling LALMs to directly learn features relevant to simultaneous translation during SFT and thereby bridge the gap between Simul-MT and Offline-MT.
To address translation errors stemming from incomplete early speech segments, we propose the Beta Decay distribution for audio truncation when constructing streaming speech data, ensuring more realistic simulation of streaming segment boundaries. We further utilize the ground-truth translation and the probability distribution from the LALM output to generate high-quality truncated speech-translation pairs.
We term this method {\bf Simul}taneous {\bf S}elf-{\bf A}ugmentation ({\bf SimulSA}) strategy, with the overall pipeline shown in Figure \ref{img:simulsa}.
Specifically, we first select a subset from the offline speech-text SFT pairs and segment the audio using the Beta Decay distribution. For each truncated speech segment, we use the probability distribution of the full ground-truth translation generated from a pre-trained LALM to speculate the most probable translation.
Finally, the LALM is mixed fine-tuned on the original SFT data and truncated speech-text pairs, allowing it to learn both offline and simultaneous translation capabilities.

We conducted extensive experiments to validate the effectiveness of our SimulSA. Results show that augmenting just {\bf 1\%} of the SFT data boosts Simul-S2TT performance by {\bf 5.1 BLEU} in the 500ms-latency scenario, effectively activating the simultaneous capabilities of base LALMs at minimal cost, without compromising offline speech translation performance. Detailed ablation study and analysis further demonstrate that employing a decay truncation distribution for audio segmentation and utilizing mix-SFT during training are crucial for performance gains. Moreover, we observe that continually scaling up augmented data yields substantial benefits under low-latency conditions. Notably, our approach requires no modifications to the model architecture and remains fully compatible with other tasks, highlighting its scalability and practical applicability.

\section{Preliminary}

\subsection{LALMs}

LALMs integrate LLM generation with audio understanding to unify the two modalities. A representative architecture, such as Qwen2-Audio \citep{chu2024qwen2}, consists of three components: an {\it Audio Encoder (AE)}, a {\it Large Language Model (LLM)}, and an {\it Adapter (ADA)} connecting them.
Given an input triplet $(\bm{u}, \bm{x}, \bm{y})$—where $\bm{u}$ is a textual prompt, $\bm{x}$ the source-language audio, and $\bm{y}$ the target-language translation—the AE encodes the audio, the ADA maps it into the LLM space, and the output is concatenated with prompt embeddings before being processed by the LLM.
The training objective is:
\begin{small}
\begin{equation}
\prod_t P_\theta(\bm{y}_t \mid \bm{y}_{\prec t}, \text{Encoder}_\phi(\bm{x}), \bm{u}),
\end{equation}
\end{small}
where $\theta$ are the LLM parameters and $\phi$ the AE and ADA parameters.

\subsection{Simul-S2TT}
Simul-S2TT incrementally receives the source speech and outputs translated text in real-time.
Given the source speech $\bm{x}_{1:s}$ with $s$ acoustic features, and the target text $\bm{y}_{1:t}$ with $t$ tokens, traditional methods use tokens in $T$ as alignment references, to find a segmentation strategy $g$ for $S$ that allows the model to learn the minimum number of audio features $g(t)$ required to translate the $t$-th token.
\section{Method}

We introduce {\bf Simul}taneous {\bf S}elf-{\bf A}ugmentation {\bf (SimulSA)} method for Simul-S2TT, to directly enhance the streaming capabilities of LALM. As shown in Figure \ref{img:simulsa}, SimulSA has three stages: 
\begin{itemize}[leftmargin=0.5cm]
    \item {\it Speech Truncation (Sec.\ref{sec:speech-truncation})}: Sample a small number of speech-text pairs $(\bm{x}_{1:s}, \bm{y}_{1:t})$ from the original offline S2TT dataset and randomly truncate the full speech \(\bm{x}_{1:s}\) to obtain \(\bm{x}_{1:s'}\);
    \item {\it Truncated Speech-to-Text Speculation (Sec.\ref{sec:speculation})}: Leverage the capabilities of LALMs to generate the translated text \(\bm{y}_{1:t'}\) corresponding to the truncated speech \(\bm{x}_{1:s'}\), thereby creating truncated speech-text pairs \((\bm{x}_{1:s'}, \bm{y}_{1:t'})\);
    \item {\it Mixed Supervised Fine-Tuning (SFT)}: Mix the full and truncated speech-text pairs obtained above, and perform a single-stage SFT on the base LALM. We also tested a two-stage SFT—first training on $\mathcal{D}$ for full translation, then on $\mathcal{D''}$ for streaming capabilities, following \citep{han2024alignsum}. However, this two-stage approach led to a significant performance drop.
\end{itemize}


\subsection{Speech Truncation}
\label{sec:speech-truncation}

Assume the SFT dataset is $\mathcal{D}$, containing $N$ speech-text pairs $\{(\bm{x}_{1:s}, \bm{y}_{1:t})\}_{i=1}^N$, we select $M$ ($M \ll N$) samples to form a subset $\mathcal{D'} \subset \mathcal{D}$.
Then, we truncate \( \bm{x}_{1:s} \) in each \((\bm{x}_{1:s}, \bm{y}_{1:t}) \in \mathcal{D'} \) to obtain truncated speech. One of the simplest ways is uniform random truncation. Assuming \( \bm{x} \) has \( s \) features, we randomly select an index \( s' \) in the interval \([1, s]\) to obtain the truncated speech \( \bm{x}_{1:s'} \).

However, considering truncated speech of all lengths uniformly presents several issues. For nearly complete speech, translation styles are similar to full translations, making extra training unnecessary. Conversely, very short truncated speech lacks enough information, preventing the model from learning effective ``wait'' signals. Additionally, accuracy in early decoding is crucial because early mistakes can accumulate and lead to irreversible consequences.

To address these issues, we first define a sub-interval \([l, r] \subset [1, s]\). This means we only sample truncation points from the interval \([l, r]\), effectively {\bf filtering out excessively short or long truncated speech}. 
Next, we consider a Beta distribution $f(X; \alpha, \beta)$ \citep{mcdonald1995generalization} with the probability density function given by:
\begin{equation}
    f(X; \alpha, \beta) = \frac{X^{\alpha-1}(1-X)^{\beta-1}}{\int_0^1 u^{\alpha-1}(1-u)^{\beta-1} \mathrm{d}u}.
\end{equation}
We set \(\alpha = 1\) and \(\beta = 3\), resulting in a monotonically decreasing function for \(f(X)\). As \(X\) increases, the gradient of \(f(X)\) diminishes, which meets our objective to {\bf prioritize early-stage translation decoding}.
Finally, since the domain of \(f(X)\) is \(X \in (0, 1)\), we construct a linear mapping from this interval to \([l, r]\). This enables us to transform \(X \sim f(X)\) into specific truncation points \(s'\). Thus, the speech truncation point \(s'\) can be expressed as:
\begin{small}
\begin{equation}
\begin{aligned}
    &~~s' = l + (r-l) \cdot x', \quad x' \sim 3(1-x)^2,\\
    \Longrightarrow &~~s' \sim f(s') = 3 \cdot \frac{(r-s)^2}{(r-l)^3},
\end{aligned}
\end{equation}
\end{small}
where $f(s')$ is the speech truncation distribution we define.




\begin{table*}[t]
\vspace{-0.05in}
\caption{BLEU and xCOMET results for different models, various chunk sizes $k$, and rollback sizes $b$. SFT size refers to the amount of training data used during the SFT phase, and $k=\infty$ refers to the offline translation.}
\centering

\renewcommand\arraystretch{0.95}
  
\setlength{\tabcolsep}{3mm}{
\resizebox{1\textwidth}{!}{
\begin{tabular}{lcccccccc|ccccccc}

\toprule
\multicolumn{16}{c}{\bf Rollback $b = 0$} \\
\midrule
\multirow{2}{*}{\bf Model} & \multirow{2}{*}{\bf SFT Size} & \multicolumn{7}{c}{\bf Chunk Size $k$ (BLEU)} & \multicolumn{7}{c}{\bf Chunk Size $k$ (xCOMET)} \\
\cline{3-16}
& & $\infty$ & $500$ & $1000$ & $1500$ & $2000$ & $3000$ & $4000$ & $\infty$ & $500$ & $1000$ & $1500$ & $2000$ & $3000$ & $4000$ \\
\midrule
Qwen2-Audio-Base & - 
& {\it 44.3} & 0.3 & 3.1 & 9.7 & 14.5 & 24.3 & 30.2 
& {\it 85.2} & 28.9 & 31.2 & 38.2 & 46.5 & 53.1 & 63.7
\\

+ SFT & 232k 
& {\it 46.1} & 0.7 & 4.8 & 12.6 & 18.2 & 24.9 & 31.5 
& {\it 87.8} & 30.1 & 33.5 & 42.7 & 51.1 & 61.8 & {\bf 69.4}
\\

\rowcolor{gray!12}
+ {\bf SFT \& SimulSA (Ours)} & 235k 
& {\it 46.0} & {\bf 7.9} & {\bf 13.4} & {\bf 20.0} & {\bf 24.3} & {\bf 29.9} & {\bf 33.8} 
& {\it 88.1} & {\bf 35.7} & {\bf 42.9} & {\bf 50.5} & {\bf 57.1} & {\bf 63.7} & 68.5
\\
\midrule

\multicolumn{16}{c}{\bf Rollback $b = 3$} \\
\midrule
\multirow{2}{*}{\bf Model} & \multirow{2}{*}{\bf SFT Size} & \multicolumn{7}{c}{\bf Chunk Size $k$ (BLEU)} & \multicolumn{7}{c}{\bf Chunk Size $k$ (xCOMET)} \\
\cline{3-16}
& & $\infty$ & $500$ & $1000$ & $1500$ & $2000$ & $3000$ & $4000$ & $\infty$ & $500$ & $1000$ & $1500$ & $2000$ & $3000$ & $4000$ \\
\midrule
Qwen2-Audio-Base & - 
& {\it 44.3} & 22.0 & 26.6 & 29.6 & 31.6 & 34.0 & 36.8 
& {\it 85.2} & 52.0 & 57.7 & 60.9 & 64.0 & 71.5 & 72.4
\\

+ SFT & 232k 
& {\it 46.1} & 29.1 & 33.2 & 35.4 & 37.0 & 38.7 & 39.3 
& {\it 87.8} & 58.4 & 66.2 & 68.9 & 70.5 & 72.5 & 74.0
\\

\rowcolor{gray!12}
+ {\bf SFT \& SimulSA (Ours)} & 235k 
& {\it 46.0} & {\bf 34.2} & {\bf 36.4} & {\bf 37.1} & {\bf 38.4} & {\bf 39.1} & {\bf 40.2} 
& {\it 88.1} & {\bf 67.3} & {\bf 69.9} & {\bf 72.9} & {\bf 73.8} & {\bf 74.5} & {\bf 75.2}
\\
\midrule

\multicolumn{16}{c}{\bf Rollback $b = 5$} \\
\midrule
\multirow{2}{*}{\bf Model} & \multirow{2}{*}{\bf SFT Size} & \multicolumn{7}{c}{\bf Chunk Size $k$ (BLEU)} & \multicolumn{7}{c}{\bf Chunk Size $k$ (xCOMET)} \\
\cline{3-16}
& & $\infty$ & $500$ & $1000$ & $1500$ & $2000$ & $3000$ & $4000$ & $\infty$ & $500$ & $1000$ & $1500$ & $2000$ & $3000$ & $4000$ \\
\midrule
Qwen2-Audio-Base & - 
& {\it 44.3} & 29.8 & 31.1 & 32.7 & 34.1 & 36.0 & 38.0 
& {\it 85.2} & 56.9 & 62.4 & 66.6 & 68.0 & 71.2 & 73.4
\\

+ SFT & 232k 
& {\it 46.1} & 37.0 & 38.6 & 39.6 & 40.3 & 41.0 & 41.2 
& {\it 87.8} & 64.9 & 68.7 & 69.6 & 71.2 & 74.5 & 74.7
\\

\rowcolor{gray!12}
+ {\bf SFT \& SimulSA (Ours)} & 235k 
& {\it 46.0} & {\bf 38.3} & {\bf 39.5} & {\bf 40.2} & {\bf 40.5} & {\bf 41.3} & {\bf 41.5} 
& {\it 88.1} & {\bf 69.4} & {\bf 71.5} & {\bf 73.3} & {\bf 74.9} & {\bf 76.9} & {\bf 77.8}
\\
\bottomrule

\end{tabular}%
}
}

\label{tab:main}%

\vspace{-0.2in}
\end{table*}

\subsection{Truncated Speech-to-Text Speculation}
\label{sec:speculation}
After obtaining the truncated speech \(\bm{x}_{1:s'}\), we need the corresponding translated text to form a truncated speech-text pair. While manual annotation is straightforward, it is time-consuming, costly, and difficult to scale. Therefore, we leverage the capabilities of foundational LALMs to automatically retrieve the translated text for \(\bm{x}_{1:s'}\).

Since LALMs generate translated text in an autoregressive manner, the translation corresponding to \(\bm{x}_{1:s'}\) must be in the form \(\bm{y}_{1:t'}\), where \(t'\) indicates the text truncation point, with \(1 \leq t' \leq t\). Therefore, determining \(\bm{y}_{1:t'}\) based on \(\bm{x}_{1:s'}\) is essential.
From the perspective of the target text, all information in \(\bm{y}_{1:t'}\) must be presented in \(\bm{x}_{1:s'}\), while maximizing the retention of translatable content. Thus, \(t'\) must meet the following conditions: (1) For each \(\bm{y}_{i} (1 \leq i \leq t')\), there must be corresponding information in \(\bm{x}_{1:s'}\); (2) For \(\bm{y}_{t'+1}\), there should be no corresponding information in \(\bm{x}_{1:s'}\).
Excluding the content already translated in \(\bm{y}_{1:t'}\) from \(\bm{x}_{1:s'}\) leaves the remaining speech information, indicating what still needs to be translated later. This forms a speech-text pair that captures the ``wait'' signal.

Building on the above analysis, we propose the Speech-to-Text Speculation technique to automatically create truncated speech-to-text pairs, drawing inspiration from Speculative Decoding \citep{leviathan2023fast}.
Specifically, we iteratively concatenate \(\bm{x}_{1:s'}\) and \(\bm{y}_{1:i} ~(i=1, 2, ..., t)\), and input them along with the prompt to the LALMs to obtain the probability distribution \(\bm{p}(\cdot | \text{prompt}, \bm{x}_{1:s'}, \bm{y}_{1:i})\) for the next token.
We believe that if the information of \(y_{i}\) has appeared in \(\bm{x}_{1:s'}\): (1) its corresponding probability should be higher than that of \texttt{<EOS>}; (2) the probability ranking of \(y_{i}\) should be among the top tokens in the vocabulary, indicating lower uncertainty from the LALMs.
Therefore, at each iteration, we determine the termination rule based on the following criteria:
\begin{small}
\begin{equation}
    \begin{aligned}
        \bm{p}(y_i) < \bm{p}(\texttt{<EOS>}) 
        ~~\bigcup~~ &\frac{1}{|\mathcal{V}|} \sum_{\text{token} \in \mathcal{V}} \mathbb{I} \left\{ \bm{p}(y_i) < \bm{p}(\text{token}) \right\} > \tau,
    \end{aligned}
    \label{eq:condition}
\end{equation}
\end{small}
where $\tau$ is the position threshold. In the \(i\)-th iteration, if the probability corresponding to \(y_i\) satisfies the Eq.\ref{eq:condition}, iteration will terminate, and \((\bm{x}_{1:s'}, \bm{y}_{1:i})\) will be the final constructed speech-text pair. Otherwise, we continue to the next iteration with \(y_{i+1}\) until termination.

\section{Experiments}

\subsection{Setup}

For the dataset, we use the CoVoST2 \citep{wang2020covost}, which contains speech and text pairs. Specifically, we use English as the source language and Chinese as the target language. The training set includes 364 hours of audio with 232,341 samples, while the test set has 25 hours of audio with 15,531 samples.

For the base model, we use Qwen2-Audio-7B \citep{chu2024qwen2} and apply SimulSA and SFT to its base version\footnote{\url{https://huggingface.co/Qwen/Qwen2-Audio-7B}}. For SFT, we utilize LoRA \citep{hu2021lora} with a rank of 8 and an alpha of 32. During training, we set the batch size to 128, the learning rate to 1e-4, and the weight decay to 0.1. We train the model using the \texttt{ms-swift} framework\footnote{\url{https://github.com/modelscope/ms-swift}} and use the checkpoint from the second training epoch for evaluation. We follow the template similar to Qwen2-Audio:
\vspace{-5pt} 

\begin{center}
\begin{tcolorbox}[colback=white,
                  colframe=black,
                  width=7.7cm,
                  arc=1mm,
                  auto outer arc,
                  boxrule=0.5pt,
                 ]
$\texttt{<audio>}\{\text{your\_speech\_file\_path}\}\texttt{</audio>}$
Detect the language and translate the speech into Mandarin: \texttt{<|en|>}
\end{tcolorbox}
\end{center}
\vspace{-5pt}

For evaluation, we test Simul-S2TT with chunk sizes \( k \) of 500, 1000, 1500, 2000, 3000, and 4000 ms. We also perform offline translation by setting \( k \) to infinity, translating after all speech is received for the best performance. In practice, we use a rollback strategy by removing the last \( b \) tokens after each step, testing \( b = 0 \), \( 3 \), and \( 5 \) (no rollback). We measure BLEU scores \citep{papineni2002bleu} using \texttt{SacreBLEU}\footnote{\url{https://github.com/mjpost/sacrebleu}} and xCOMET scores \citep{guerreiro2024xcomet} using \texttt{XCOMET-XXL}\footnote{\url{https://huggingface.co/Unbabel/XCOMET-XXL}}.

For hyper-parameters, the sampling interval \([l, r]\) in speech truncation is \( l = 500ms \) and \( r = 5000ms \), or the maximum feature count if needed.
The position threshold \( \tau \) is \( \tau = 100 / v \), where \( v = 151{,}646 \), so \( \tau = 6.6 \times 10^{-4} \).
The augmented data size \( m \) is set to 3,000. An ablation study on this is in Section \ref{sec:augsize}.



\subsection{Main Results}

Table \ref{tab:main} presents the main results of each model under varying inference settings. Our analysis yields the following findings:

\textbf{Low-Cost for High-Improvement.}
Augmenting SFT with SimulSA increases the corpus only slightly (232k → 235k) yet yields substantial BLEU gains. For a fixed chunk size, improvements are largest at smaller rollback sizes, which use fewer inference tokens. For example, with chunk size 1500, BLEU improves over SFT by +7.4 (rollback 0), +1.7 (rollback 3), and +0.6 (rollback 5). This trend holds across chunk sizes, showing that {\it a small amount of SimulSA data plus our training strategy markedly improves simultaneous translation with minimal additional cost.}

\textbf{Low-Latency Benefit Most.}
As chunk size decreases (lower latency), gains grow. At rollback 0, SimulSA+SFT improves BLEU by +7.2, +8.6, and +7.4 for chunk sizes 500, 1000, and 1500, versus +5.0 and +2.3 for chunk sizes 3000 and 4000. This underscores the effectiveness of our speech-truncation strategy (Sec.~\ref{sec:speech-truncation}) in low-latency regimes: {\it gains are modest at large chunks (high latency) but substantial at small chunks.}


\textbf{Robust Offline MT.}
Although simultaneous and offline MT distributions differ (Table \ref{img:simul-case}), adding SimulSA does not harm offline MT: at \(k=\infty\), BLEU changes within random variation. Hence, {\it SimulSA improves simultaneous MT without compromising offline MT, ensuring robustness and reliability for real-world applications.}
\section{Ablation and Analysis}

\begin{figure}[t]
  \centering
  \includegraphics[width=0.85\columnwidth]{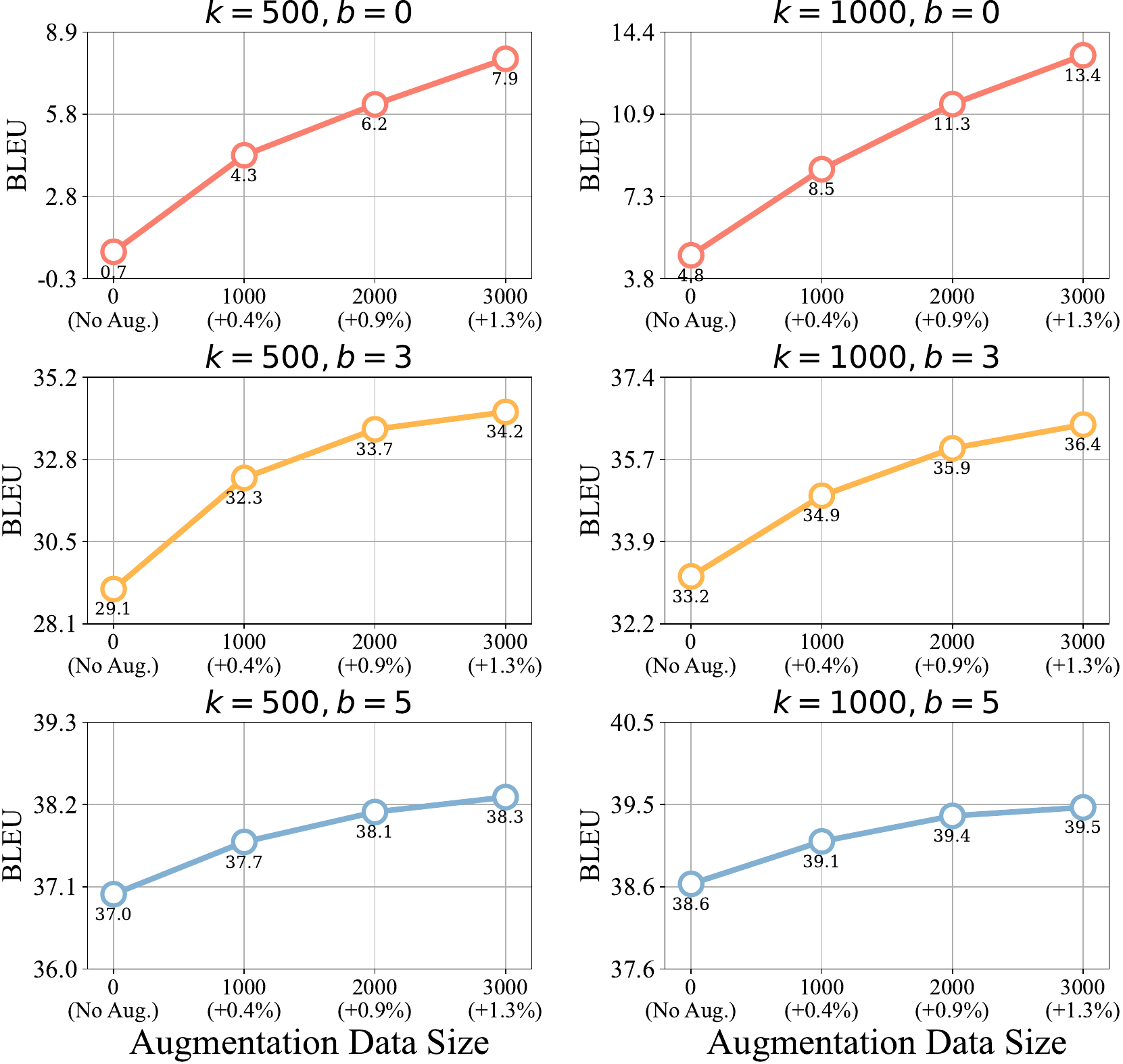}  
  \vspace{-0.1in}
  \caption{Ablation of self-augmentation data size for different $k$ and $b$.}
  \label{img:augsize}
\vspace{-0.15in}
\end{figure}

\subsection{Self-Augmentation Data Size}
\label{sec:augsize}

To thoroughly investigate how the size of SimulSA data ($m$) affects performance, we conduct an ablation study using $m \in \{1000, 2000, 3000\}$, performing SFT for each case. BLEU scores were measured for each $m$ under the settings of
\[
\{k | k=500, 1000, 1500, 2000\} \times \{b|b=0,3,5\}.
\]
As shown in Figure \ref{img:augsize}, increasing the amount of augmented data consistently boosts BLEU scores across all $k$ and $b$ values. When $b = 0$ (no rollback), BLEU improvements scale almost linearly with additional data, suggesting further gains as more augmented data is added. For $b \geq 3$ (higher $b$ means greater inference latency), BLEU improvements plateau once the augmented data reaches about 1.3\% of the original dataset. This saturation indicates that only a modest increase in augmented data is needed to meaningfully enhance the performance of these stronger baselines.

\begin{figure}[t]
  \centering
  \includegraphics[width=1\columnwidth]{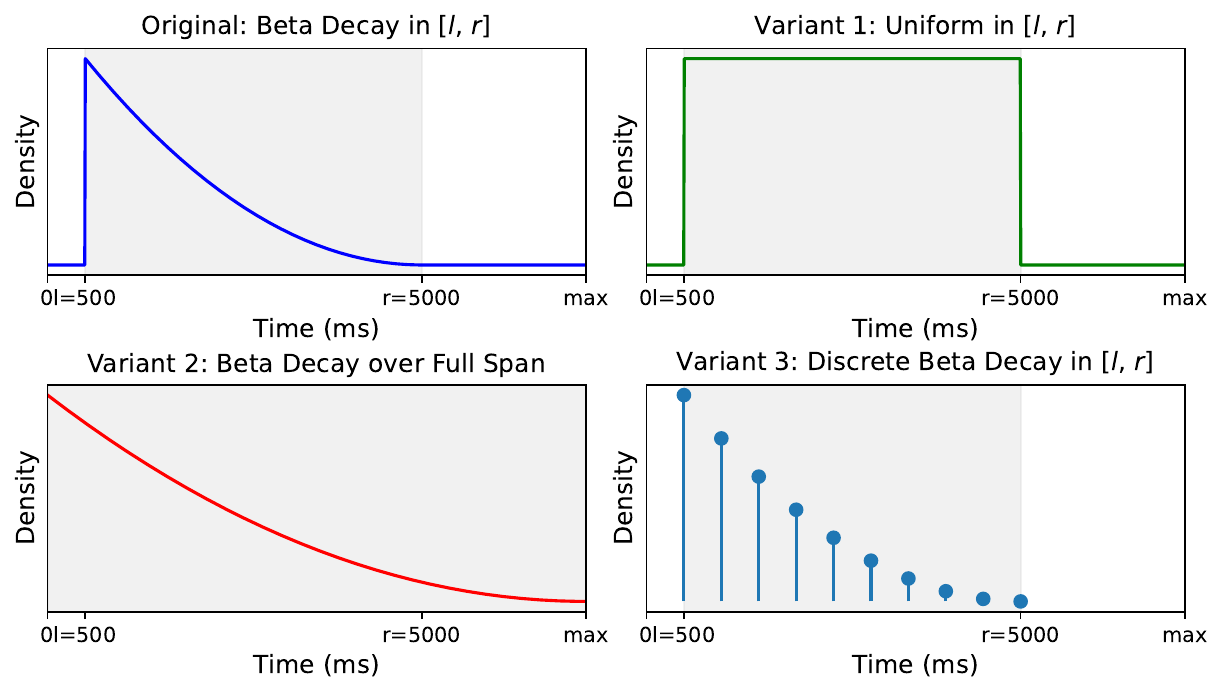}  
  \vspace{-0.25in}

  \caption{Variants of speech truncation distribution: {\bf Original}—continuous beta decay within \([l, r]\); {\bf Variant 1}—continuous uniform within \([l, r]\); \textbf{Variant 2}—continuous beta decay over the full speech span; \textbf{Variant 3}—discrete beta decay every 500ms within \([l, r]\). The shaded area indicates the truncation sampling interval.}

  \label{img:dist}
\vspace{-0.1in}
\end{figure}

\begin{table}[h]
\caption{BLEU scores for Original and Variant 1–3 truncation distributions across speech chunk sizes (\(k=500, 1000, 1500, 2000\)) and rollback values (0, 3, 5).
}
\centering
\renewcommand\arraystretch{1.2}
  
\setlength{\tabcolsep}{1mm}{
\resizebox{1\columnwidth}{!}{
\begin{tabular}{lcccc}

\toprule

\multirow{2}{*}{\shortstack{{\bf Truncation}\\{\bf Distribution}}} & \multicolumn{4}{c}{\bf Chunk Size $k$} \\
\cline{2-5}
& $500$ & $1000$ & $1500$ & $2000$ \\
\midrule

Original & 7.9 / 34.2 / 38.3 & 13.4 / 36.4 / 39.5 & 20.0 / 37.1 / 40.2 & 24.3 / 38.4 / 40.5 \\
Variant 1 & 7.0 / 33.5 / 38.1 & 12.2 / 36.6 / 39.8 & 19.2 / 37.0 / 40.0 & 23.4 / 38.2 / 40.5 \\
Variant 2 & 4.3 / 31.5 / 37.6 & 8.2 / 34.9 / 38.9 & 17.3 / 36.7 / 39.8 & 21.8 / 37.6 / 40.4 \\
Variant 3 & 6.5 / 33.1 / 37.4 & 10.9 / 35.7 / 39.0 & 18.8 / 36.9 / 40.0 & 22.6 / 38.1 / 40.5 \\

\bottomrule

\end{tabular}%
}
}

\label{tab:dist-results}%

\vspace{-0.1in}
\end{table}

\subsection{Truncation Distribution Design}
In Section \ref{sec:speech-truncation}, we proposed a decaying sampling distribution over speech truncation points to avoid extremely short or long segments and to emphasize learning from early speech–text prefixes. We evaluate this design via an ablation on the truncation distribution.
We compare three variants of the sampling distribution (Figure \ref{img:dist}) and report BLEU in Table \ref{tab:dist-results}. The results show:
\textbf{Variant 2} performs worst, likely because it oversamples short, uninformative or noisy prefixes and also yields very long segments that approximate offline translation;
\textbf{Variant 3} underperforms due to reduced sampling diversity, which weakens robustness and degrades translation quality;
\textbf{Variant 1} improves by avoiding extreme segment lengths and increasing truncation diversity, but still fails to sufficiently strengthen learning on early simultaneous-translation segments.
By contrast, our beta-decay distribution yields the best BLEU. It preserves diversity within a bounded interval while biasing toward early truncation points. The gain is especially pronounced when rollback is disabled, indicating reduced error accumulation from early mistranslations.

\section{Conclusion}

We address the challenge of equipping large audio--language models (LALMs) for real-time speech translation. We present SimulSA, a lightweight augmentation that derives truncated audio-text segments from full pairs. Adding only 1\% augmented data during training enables a single model to support both streaming and offline translation without architectural changes. SimulSA yields substantial gains under low-latency constraints and works well with other techniques, enhancing the practicality of LALMs for real-time deployment.

\section{Acknowledgment}
This work was supported in part by the Science and Technology Development Fund of Macau SAR (Grant Nos. 0007/2024/AKP, FDCT/0070/2022/AMJ, FDCT/060/2022/AFJ), and the UM and UMDF (Grant Nos. MYRG-GRG2023-00006-FST-UMDF, MYRG-GRG2024-00165-FST-UMDF, EF2024-00185-FST, EF2023-00151-FST, EF2023-00090-FST).




\bibliographystyle{unsrtnat}
\bibliography{strings,refs}

@article{cho2016can,
  title={Can neural machine translation do simultaneous translation?},
  author={Cho, K},
  journal={arXiv Preprint, CoRR, arXiv: abs/1606.02012},
  year={2016}
}

@inproceedings{raffel2017online,
  title={Online and linear-time attention by enforcing monotonic alignments},
  author={Raffel, Colin and Luong, Minh-Thang and others},
  booktitle={International conference on machine learning},
  pages={2837--2846},
  year={2017},
  organization={PMLR}
}

@inproceedings{arivazhagan2019monotonic,
  title={Monotonic Infinite Lookback Attention for Simultaneous Machine Translation},
  author={Arivazhagan, Naveen and Cherry, Colin and others},
  booktitle={Proceedings of the 57th Annual Meeting of the Association for Computational Linguistics},
  pages={1313--1323},
  year={2019}
}

@article{guo2024sillm,
  title={SiLLM: Large Language Models for Simultaneous Machine Translation},
  author={Guo, Shoutao and Zhang, Shaolei and others},
  journal={arXiv preprint arXiv:2402.13036},
  year={2024}
}

@inproceedings{iranzo2020direct,
  title={Direct segmentation models for streaming speech translation},
  author={Iranzo-S{\'a}nchez, Javier and Gim{\'e}nez Pastor, Adri{\'a}n and others},
  booktitle={Proceedings of the 2020 Conference on Empirical Methods in Natural Language Processing (EMNLP)},
  pages={2599--2611},
  year={2020},
  organization={Association for Computational Linguistics}
}

@inproceedings{chen2021direct,
  title={Direct Simultaneous Speech-to-Text Translation Assisted by Synchronized Streaming ASR},
  author={Chen, Junkun and Ma, Mingbo and others},
  booktitle={Findings of the Association for Computational Linguistics: ACL-IJCNLP 2021},
  pages={4618--4624},
  year={2021}
}

@inproceedings{liu2024recent,
  title={Recent advances in end-to-end simultaneous speech translation},
  author={Liu, Xiaoqian and Hu, Guoqiang and others},
  booktitle={Proceedings of the Thirty-Third International Joint Conference on Artificial Intelligence},
  pages={8142--8150},
  year={2024}
}

@inproceedings{dong2022learning,
  title={Learning When to Translate for Streaming Speech},
  author={Dong, Qian and Zhu, Yaoming and others},
  booktitle={Proceedings of the 60th Annual Meeting of the Association for Computational Linguistics (Volume 1: Long Papers)},
  pages={680--694},
  year={2022}
}

@inproceedings{zhang2023end,
  title={End-to-End Simultaneous Speech Translation with Differentiable Segmentation},
  author={Zhang, Shaolei and Feng, Yang},
  booktitle={Findings of the Association for Computational Linguistics: ACL 2023},
  pages={7659--7680},
  year={2023}
}

@article{zaidi2021decision,
  title={Decision attentive regularization to improve simultaneous speech translation systems},
  author={Zaidi, Mohd Abbas and Lee, Beomseok and others},
  journal={arXiv preprint arXiv:2110.15729},
  year={2021}
}

@inproceedings{liu2021cross,
  title={Cross attention augmented transducer networks for simultaneous translation},
  author={Liu, Dan and Du, Mengge and others},
  booktitle={Proceedings of the 2021 Conference on Empirical Methods in Natural Language Processing},
  pages={39--55},
  year={2021}
}

@article{polak2023incremental,
  title={Incremental blockwise beam search for simultaneous speech translation with controllable quality-latency tradeoff},
  author={Pol{\'a}k, Peter and Yan, Brian and others},
  journal={arXiv preprint arXiv:2309.11379},
  year={2023}
}

@inproceedings{papi2023attention,
  title={Attention as a Guide for Simultaneous Speech Translation},
  author={Papi, Sara and Negri, Matteo and Turchi, Marco},
  booktitle={Proceedings of the 61st Annual Meeting of the Association for Computational Linguistics (Volume 1: Long Papers)},
  pages={13340--13356},
  year={2023}
}

@article{wang2020covost,
  title={Covost 2 and massively multilingual speech-to-text translation},
  author={Wang, Changhan and Wu, Anne and Pino, Juan},
  journal={arXiv preprint arXiv:2007.10310},
  year={2020}
}

@article{chu2023qwen,
  title={Qwen-audio: Advancing universal audio understanding via unified large-scale audio-language models},
  author={Chu, Yunfei and Xu, Jin and others},
  journal={arXiv preprint arXiv:2311.07919},
  year={2023}
}

@article{chu2024qwen2,
  title={Qwen2-audio technical report},
  author={Chu, Yunfei and Xu, Jin and others},
  journal={arXiv preprint arXiv:2407.10759},
  year={2024}
}

@article{das2024speechverse,
  title={Speechverse: A large-scale generalizable audio language model},
  author={Das, Nilaksh and Dingliwal, Saket and others},
  journal={arXiv preprint arXiv:2405.08295},
  year={2024}
}

@article{tang2023salmonn,
  title={Salmonn: Towards generic hearing abilities for large language models},
  author={Tang, Changli and Yu, Wenyi and others},
  journal={arXiv preprint arXiv:2310.13289},
  year={2023}
}

@article{hu2021lora,
  title={Lora: Low-rank adaptation of large language models},
  author={Hu, Edward J and Shen, Yelong and Wallis, Phillip and Allen-Zhu, Zeyuan and Li, Yuanzhi and Wang, Shean and Wang, Lu and Chen, Weizhu},
  journal={arXiv preprint arXiv:2106.09685},
  year={2021}
}

@inproceedings{papineni2002bleu,
  title={Bleu: a method for automatic evaluation of machine translation},
  author={Papineni, Kishore and Roukos, Salim and Ward, Todd and Zhu, Wei-Jing},
  booktitle={Proceedings of the 40th annual meeting of the Association for Computational Linguistics},
  pages={311--318},
  year={2002}
}

@article{mcdonald1995generalization,
  title={A generalization of the beta distribution with applications},
  author={McDonald, James B and Xu, Yexiao J},
  journal={Journal of Econometrics},
  volume={66},
  number={1-2},
  pages={133--152},
  year={1995},
  publisher={Elsevier}
}

@inproceedings{leviathan2023fast,
  title={Fast inference from transformers via speculative decoding},
  author={Leviathan, Yaniv and Kalman, Matan and Matias, Yossi},
  booktitle={International Conference on Machine Learning},
  pages={19274--19286},
  year={2023},
  organization={PMLR}
}

@inproceedings{han2024alignsum,
  title={AlignSum: Data Pyramid Hierarchical Fine-tuning for Aligning with Human Summarization Preference},
  author={Han, Yang and Wang, Yiming and Wang, Rui and Chen, Lu and Yu, Kai},
  booktitle={Findings of the Association for Computational Linguistics: EMNLP 2024},
  pages={8506--8522},
  year={2024}
}

@article{guerreiro2024xcomet,
  title={xcomet: Transparent machine translation evaluation through fine-grained error detection},
  author={Guerreiro, Nuno M and Rei, Ricardo and Stigt, Daan van and Coheur, Luisa and Colombo, Pierre and Martins, Andr{\'e} FT},
  journal={Transactions of the Association for Computational Linguistics},
  volume={12},
  pages={979--995},
  year={2024},
  publisher={MIT Press 255 Main Street, 9th Floor, Cambridge, Massachusetts 02142, USA~…}
}

\end{document}